\documentstyle[epsf,rotate]{article}

\begin{document}

\begin{flushright}
Liverpool Preprint: LTH 629\\
 \end{flushright}

%%\draft  % \draft command makes pacs numbers print

\vspace{5mm}
\begin{center}
{\LARGE \bf An unquenched lattice QCD calculation of the mass of 
the bottom quark }\\[10mm] 
{\large\it UKQCD Collaboration}\\[3mm]
 
 {\bf C. McNeile, C.~Michael, and Gavin Thompson \\
Theoretical Physics Division, Dept. of Mathematical Sciences, 
          University of Liverpool, Liverpool L69 3BX, UK 
 }\\[2mm]

\end{center}

\begin{abstract}
We compute the $b$  quark mass from  dynamical lattice QCD with clover
quarks. The calculation is done at a fixed lattice 
spacing with sea quark masses as low as 
half the strange quark mass.
Our final result is  $\overline{m_b} (\overline{m_b} )$ =
4.25(2)(11) GeV, where the first error is statistical and 
the last error is the systematic uncertainty.
 \end{abstract}
% Lattice QCD Calculations
%%\pacs{12.38.Gc} 

% body of paper here

\section{Introduction}

The mass of the bottom quark is a fundamental parameter
of the standard model~\cite{Lubicz:2000ch}. 
To extract the $b$ mass from experiment,
QCD corrections must be computed reliably. The best
way to do this is use lattice QCD.
The different methods of computing the mass of the 
bottom quark have recently been reviewed by
El-Khadra and Luke~\cite{El-Khadra:2002wp}.

The particle data table quotes the mass of the b quark
in the $\overline{MS}$ scheme at the b mass to lie between 4.0 and
4.5 GeV~\cite{Hagiwara:2002fs}. 
It is particularly important to reduce the error
on the $b$ quark mass, because it is the cause of the 
largest uncertainty on the determination of $V_{ub}$
from the total inclusive $B$ meson decay $b \rightarrow ul\nu$.
El-Khadra and Luke~\cite{El-Khadra:2002wp} note that a 100 MeV
error on $m_b$ corresponds to a 6 \% error on the $V_{ub}$ 
CKM matrix element (currently only known to
19 \% accuracy~\cite{Hagiwara:2002fs}).

In this paper we use unquenched lattice QCD
with static-light mesons to extract $m_b$.
As we discuss in section~\ref{eq:system},
the error due to the use 
of static (leading order
HQET) approximation
is only of order 30 MeV~\cite{Gimenez:2000cj},
hence the static limit has
an important role for the phenomenology of
determining $V_{ub}$.

\section{Details of lattice calculations}

We used non-perturbatively improved clover fermions
in both the sea and valence quarks. The Wilson gauge action
is used for the gluons.
The full details of the 
actions and details of the unquenched calculation are
described in ~\cite{Allton:1998gi,Allton:2001sk,Allton:2004qq}.
%%The parameters of the lattice calculations are in 
%%table~\ref{tab:bareRESULTS}.

We use static quarks for the heavy mesons. The lattice binding energy is
extracted from  a two point correlator using variational smearing techniques.
The local two point function is
 %%%%
\begin{eqnarray}
C(t) & = &  \sum_{x} 
\langle 0 \mid 
\Phi_B(x,t) \Phi_B^\dagger(x,0) 
\mid 0 \rangle \\
     & = & Z^2 \exp( - a{ \cal E } t)
\end{eqnarray}
where $\Phi_B$ is the interpolating operator for  static-light mesons.
We have already published~\cite{Green:2003zz} an extensive analysis of
the spectrum of static-light mesons. Our previous
paper~\cite{Green:2003zz} also describes the all-to-all  propagators
used to improve the statistical accuracy and the fuzzing methods used.

In table~\ref{tab:bareRESULTS} we present our results  for the lattice
binding energy.
All the data sets used $\beta$ = 5.2.
Data sets $DF1$ and $DF2$ used a clover coefficient of 
1.76, while all the others used the non-perturbative value of 
2.0171.
The results for the data sets: DF1, DF2 have already
been published~\cite{Green:2003zz}. The ensemble size for data sets DF4
and DF5 have been trebled over the results previously 
published~\cite{Green:2003zz}. 
The data from ensembles DF5 and DF6 are new.

\begin{table}[tb]
\begin{tabular}{cc|cc|cc|||c|c}
Name & 
No.
& $r_0 m_{PS}$ & $\kappa_{sea}$ & $\kappa_{val}$ & 
Volume  &  ${a \cal E}$ &  $\Lambda_{static}$ GeV \\ \hline
%%%
DF1 & 20 & 1.92(4) & 0.1395   & 0.1395 &
$12^3 \; 24$  & 0.87(1) &  0.59(5)  \\ 
%%%
DF2 & 78 & 1.94(3) & 0.1395   & 0.1395 &
$16^3 \; 24$  & 0.842(5) & 0.55(2) \\ 
%%%
DF3 & 60 & 1.93(3) & 0.1350 & 0.1350 &
$16^3 \; 32$  & $0.772^{+7}_{-8}$ & 0.69(7) \\ 
%%%
DF4 & 60 & 1.48(3) & 0.1355 & 0.1355 & 
$16^3 \; 32$  & $0.739^{+9}_{-8}$ & 0.66(5) \\ 
%%%%
DF5 & 60 & 1.82(3) & 0.1355 & 0.1350 & 
$16^3 \; 32$  & $0.748^{+9}_{-8}$ & 0.68(5) \\ 
%%%%
DF6 & 55 & 1.06(3) & 0.1358 & 0.1358 & 
$16^3 \; 32$  & $0.707^{+14}_{-12}$ & 0.64(7) \\ 
  \end{tabular}
%%% draper_magic  \vsbtc
  \caption{
 Lattice binding energy (${a \cal E}$)
and physical binding energy ($\Lambda_{static}$) for each data set.
}
\label{tab:bareRESULTS}
\end{table}

\section{Extracting the quark mass} \label{se:BasicIdea}

 We evaluate the mass of a pseudoscalar heavy-light meson from lattice
QCD  with static heavy quark and compare with the experimental mass
value. This gives information about the $b$-quark mass.  The strange
quark mass is accessible in lattice evaluations, so to minimise
extrapolation, we use the $B_s$ meson for this comparison. We still need
to extrapolate the sea quark  mass to the experimental value and we
discuss this later.

 In this section we will describe the central  values for our
calculation. We discuss systematic uncertainties in
section~\ref{eq:system}. The quantity ${\cal E}$, from the lattice
calculation, contains an unphysical  $\frac{1}{a}$ divergence ($\delta
m$) that must be subtracted off to obtain the physical binding 
energy ($\Lambda_{static}$).
\begin{equation}
\Lambda_{static} = {\cal E } - \delta m
\end{equation}
%%%
 The pole quark mass is determined from
\begin{equation}
m_b^{pole} = M_{B_s} - \Lambda_{static}
\end{equation}
The physical value~\cite{Hagiwara:2002fs} of the meson mass
$M_{B_s}$  (5.369 GeV) is used.

In the static theory  $\delta m$ has been
calculated to two loops by 
Martinelli and Sachrajda~\cite{Martinelli:1998vt}.
%%%
%%
\begin{eqnarray}
a \delta m & = & 2.1173 \alpha_s( \overline{m_b}) + 
\{ ( 3.707 - 0.225 n_f ) \log(\overline{m_b}a)  \nonumber \\
& - & 1.306 - n_f ( 0.104 + 0.1 c_{SW} - 0.403 c_{SW}^2) \}
\alpha_s( \overline{m_b})^2
\label{eq:deltaMLOw}
\end{eqnarray}
%%%
%%%
where $n_f$ is the number of sea quarks and $c_{SW}$ is 
the coefficient of the clover term.
We discuss estimates of the next order to $a \delta m$
in section~\ref{eq:system}.

The pole mass (see Kronfeld for a review~\cite{Kronfeld:1998di})  is
converted to $\overline{MS}$ using continuum perturbation 
theory~\cite{Gray:1990yh}.
 %%%%
\begin{equation}
m_b^{\overline{MS}}(\mu) = 
Z_{pm}(\mu)
m_b^{pole}
 + O(1/m_b)
\label{eq:masterEq}
\end{equation}
%%%%
where
%%%
%%%%  check additional bracket.
%%%%%
\begin{equation}
Z_{pm}(\mu=m_b) = 1  -\frac{4}{3} 
\frac{\alpha_s( \overline{m_b})  }{ \pi}
-
(11.66 - 1.04 n_f) ( \frac{\alpha_s( \overline{m_b})  }{ \pi} )^2
\label{eq:ZPMlow}
\end{equation}
%%%
The perturbative correction between the pole mass and
the $\overline{MS}$ mass is known to 
$O(\alpha^3)$~\cite{Melnikov:2000qh,Chetyrkin:1999qi}.
The perturbative series connecting the pole mass with the
$\overline{MS}$ mass is badly behaved due to
renormalons (see~\cite{Beneke:1998ui} for a review).
The lattice matching is only done to $O(\alpha^2)$, hence we convert
the pole mass to $\overline{MS}$ at the same order, using a consistent
coupling, so the differences in the series are physical.

We use the values of the coupling using the values of  $\Lambda_{QCD}$ from
the joint UKQCD and QCDSF paper~\cite{Booth:2001qp}. The four loop 
expression for $\alpha_s$ is used~\cite{Chetyrkin:2000yt} to determine
the coupling from $\Lambda_{QCD}$. We consistently use $n_f=2$ in all
the perturbative expressions. For $\kappa_{sea} = 0.1355$ (0.1350) we use
$\Lambda_{QCD}$ = $0.178$ ( $0.173$) MeV~\cite{Booth:2001qp}. 
We use the same value
of $\Lambda_{QCD}$ for the two data sets $DF1$ and $DF2$, where
$\Lambda_{QCD}$ has not been computed.

In~\cite{Green:2003zz} we estimated the mass of the strange quark using
the pseudoscalar made out of strange quarks~\cite{McNeile:2000hf}. This
provided $r_0 m_{PS} \equiv 1.84$. This value is close to $r_0 m_{PS} =
1.82^{+3}_{-1}$ for $DF5$ data set~\cite{Allton:2001sk} hence we use the
binding energy from that data set as the value at strange.  
This is  a partially quenched analysis.
The data
sets $DF1$ and $DF2$  also have a  sea quark mass close to the  strange
quark mass~\cite{McNeile:2000hf}.

To determine the lattice spacing we use the measured value for $r_0/a$
from the potential with the `physical' value of $r_0$ as 0.525(25)
fm~\cite{Dougall:2003hv}. We discuss in more detail the systematic error
from the choice of  $r_0$ in section~\ref{eq:system}. 
Hence our best estimate of $m_b(m_b)$ = 
$4.25(2)$ GeV from the $DF5$ data set, where the errors are
statistical only.

\section{Computing the systematic uncertainties} \label{eq:system}

Gimenez et al.~\cite{Gimenez:2000cj} discuss the systematic error
from the neglect of the $1/m_b$ terms in
the static limit.
Heavy quark effective field theory parametrises
the heavy mass corrections to the 
mass of a heavy-light meson $M_B$.
 \begin{equation}
M_B = m_b  + \Lambda_{static} - 
\frac{\lambda_1} {2 m_b} - \frac{ 3  \lambda_2} {2 m_b}
 \end{equation}
 where $\Lambda_{static}$ is the static binding energy,
$\lambda_1$ is the matrix element due to the 
insertion of the kinetic energy and $\lambda_2$
is the matrix element due to the insertion of the  chromomagnetic
operator.
The value of $\lambda_2 \sim 0.12 \; {\mathrm GeV}^2$ can be obtained from
the experimental mass splitting between the $B^{\star}$
and $B$ mesons.  The value of $\lambda_1$ is much harder
to estimate.
Gimenez et al.~\cite{Gimenez:2000cj} use a 
range of $\lambda_1$ from  -0.5 to 0.0 in  GeV$^2$. This includes
the determination from quenched lattice QCD of
$\lambda_1 = -(0.45 \pm 0.12) {\mathrm GeV}^2$ by
Kronfeld and Simone~\cite{Kronfeld:2000gk}.
JLQCD have recently tried to compute $\lambda_1$
using NRQCD~\cite{Aoki:2003jf}. 
As suggested by Gimenez et al.~\cite{Gimenez:2000cj},
using a symmetric error
of 30 MeV to parameterise the neglected $1/m_b$
terms seems reasonable to us.

The specification of the strange quark mass, described in 
section~\ref{se:BasicIdea}, essentially relies on
the  physical K mass. Since the $\phi$ meson has a very narrow width, it
may also be used to specify the  strange quark mass. In their study
using essentially the same lattice parameters, JLQCD~\cite{Aoki:2002uc} see
approximately a 10\% difference between using the $\phi$ and the $K$ to
set the strange quark mass. Motivated by JLQCD's result, we use a
symmetric error of 5\%  as an estimate of the additional uncertainty in
our estimate of the  strange quark mass. This induces an error of
60 MeV in $m_b(m_b)$ for the $DF5$ data set.

The data sets $DF1$, $DF2$ were generated using a different value of
$c_{SW}$ to that used to generate data set $DF5$, hence they can not be
used to estimate the size of the lattice spacing effects. The comparison
of the results between data set $DF1$ and $DF2$ can in principle 
be used to estimate
finite size effects. As the physical size of a size of the lattice
changes from 1.83 fm (DF1) to  2.44 fm (DF2), $m_b(m_b)$ changes
from 4.33(2) to 4.369(7) GeV.
Hence, a simple estimate of the finite size effects
in date set $DF5$ (size of box 1.77fm) is -39 MeV.
In quenched QCD Duncan et al.~\cite{Duncan:1994in} found
no finite size effects in the binding energy for lattice lengths: 
1.3, 1.8 and 2.2 fm. Although, finite size effects in 
quenched and unquenched QCD can be very different, we think
it more likely that the differences in two data sets is
due to a statistical fluctuation on the smaller 
lattice. 
This is supported by the fact that
UKQCD saw no finite size effects 
in the light hadron spectrum between DF1 
and DF2~\cite{Allton:1998gi}.

The choice of coupling is a systematic error.
Gimenez et al.~\cite{Gimenez:2000cj} use $\Lambda_{QCD}^{n_f=2}$
= 300 MeV as the central value. We  use the 
result for $\Lambda_{QCD}$ determined from 
the $DF4$ and $DF5$ data sets~\cite{Booth:2001qp}. 
We don't feel that it is appropriate to use 
the values of $\Lambda_{QCD}$ from experiment 
(as done by Gimenez et al.~\cite{Gimenez:2000cj})
even for deriving a systematic error.
The agreement between  $\Lambda_{QCD}$
from lattice QCD calculations and experiment
is not good for calculations that use clover fermions
for the sea quarks~\cite{Aoki:2002uc,Kaneko:2001ux,Booth:2001qp}.
We assume that the discrepancy will be reduced as calculations are 
done with lighter
sea quark masses and finer lattice spacings.

To estimate the effect of the chiral extrapolation of
the sea quark masses, we extrapolated $\Lambda_{static}$
from data sets $DF4$ and $DF5$ linearly in $(r_0 m_{PS})^2$
to $r_0 m_{PS}$ = 1.93 (the same as for data set $DF3$).
The extrapolated result for $\Lambda_{static}$ at 
$r_0 m_{PS}$ = 1.93 at $\kappa_{sea}$ = 0.1355 was consistent 
with $\Lambda_{static}$ on data set DF3 ($\kappa_{sea}$ = 0.1350).
We see no evidence for the dependence of
$\Lambda_{static}$ on the sea quark mass.
Gimenez et al.~\cite{Gimenez:2000cj} see a slight
increase in the lattice binding energy with decreasing quark mass.
There are potentially non-analytic $m_{PS}^3$ terms in the 
mass dependence of the binding energy~\cite{Goity:1992tp}.

To estimate the systematic errors on the perturbative matching we did 
a number of things. Following Gimenez et al.~\cite{Gimenez:2000cj}
we compared taking the product of the two
perturbative factors in equation~\ref{eq:masterEq} against
expanding the perturbative expressions and only keeping 
$O(\alpha_s^2)$ terms. This increases the mass for data set
$DF5$ by 25 MeV.

In quenched QCD the next order correction to 
equation~\ref{eq:deltaMLOw} has been computed numerically
by two groups using  different 
techniques~\cite{DiRenzo:2000nd,Trottier:2001vj}.
As the two groups obtained essentially the same
result we will use the result of 
Di Renzo and Scorzato~\cite{DiRenzo:2000nd}.
In quenched QCD the next order correction to 
equation~\ref{eq:deltaMLOw} is~\cite{Gimenez:2000cj}
 \begin{equation}
a \delta m^{(3)} = 
( \overline{X_2}
+ 6.48945  ( -3.57877 + \log(\overline{m_b}a) ) 
  ( 3.29596 + \log(\overline{m_b}a)  )
)  \alpha_s(\overline{m_b})^3
 \label{eq:delMNext}
 \end{equation}
%%%
 where $\overline{X_2}$ is the number from 
the numerical calculation.
Di Renzo and Scorzato~\cite{DiRenzo:2000nd}
obtain $\overline{X_2}$ = 86.2(0.6)(1.0)
for quenched QCD. Di Renzo and Scorzato
have computed $X_2$ for $n_f=2$
Wilson fermions~\cite{DiRenzo:2004xn}.
The new result also involves a lattice calculation of 
the $\overline{MS}$ coupling for Wilson fermions, 
so it is not obvious
how to incorporate the new result into this analysis.

The next order to the connection between the pole mass
and $\overline{MS}$ mass is known in the 
continuum (equation~\ref{eq:ZPMlow}).
\begin{equation}
Z_{pm}^{(3)} =
-  (157.116 - 23.8779 n_f  + 0.6527 n_f^2) 
(\frac{\alpha_s( \overline{m_b})}{\pi})^3
\label{eq:ZPMnextOrder}
\end{equation}
Because equation~\ref{eq:delMNext} is  known for
quenched QCD, we do not use it for
our central result. It does seem appropriate to use
equation~\ref{eq:delMNext} and equation~\ref{eq:ZPMnextOrder}
to estimate the systematic errors due to the neglect
of higher order terms. We set $n_f$ = 0 in 
equation~\ref{eq:ZPMnextOrder}.
Adding in the next order term reduces 
the mass of the bottom quark by 12 MeV for data set $DF5$.

Because of limitations in computer time, the unquenched calculations
are at fixed lattice spacing (so the continuum limit has not been
taken) and fairly heavy sea quark masses are used. This means that
different lattice quantities produce slightly different values of the
lattice spacing.  In~\cite{Dougall:2003hv} the values of $r_0$ from
various calculations that used clover fermions are collected together.
The results were in the range $r_0$ = 0.5 to 0.55 fm.  This motivates
our choice of $r_0$ = 0.525(25) fm.  Using mass splittings in Upsilon
on improved staggered configurations with measurements of the
potential MILC~\cite{Aubin:2004wf} quote $r_0$ = 0.467 fm. This is
$2.3 \sigma$ from our central value for $r_0$. In the
graph~\cite{Davies:2003ik} of spread of variation of lattice spacings
from different physical quantities, the most dramatic failures of the
quenched approximation occur for $P-S$ and $2S -1S$ mass splittings in
Upsilon and the pion decay constant.  We speculate that these
quantities are more sensitive to the heavy quark potential at the origin that
depends on $n_f$ from the running of the coupling 
(see~\cite{Bernard:2000gd} 
for a discussion of this).  
The relatively
strong dependence of the Upsilon mass splittings and pion decay
constant on the sea quark mass and lattice spacing doesn't make them a
good choice to set the scale for current unquenched calculations with
clover fermions.  Hence we feel that using the value from MILC for
$r_0$ (as advocated in~\cite{Wingate:2003gm}) will artificially
inflate the error bars, so we stick to our original estimate of $r_0$
= 0.525(25) fm.

The  perturbative analysis,
used in the this section,
assumes that the sea quark mass
is zero. However, the sea quark masses used in current calculations
with Wilson-like quarks are not negligible. At $\kappa_{sea}$ = 0.1355
and 0.1350, the vector definitions of the sea quark mass 
(in units of the lattice spacing)
are
0.026 and 0.044 respectively.
The light quark mass dependence has been computed by
      Bali and Boyle~\cite{Bali:2002wf}.

For light quark masses below 0.1 (in lattice units) Bali and 
Boyle~\cite{Bali:2002wf} provide a quadratic parameterisation of
the light quark mass dependence of $\delta m$. The expression 
in equation~\ref{eq:deltaMLOw} gets modified to
%%%
%%%
\begin{eqnarray}
a \delta m & = & 2.1171 \alpha_s( \overline{m_b}) + 
( ( 3.707 - 0.225 n_f ) \log(\overline{m_b}a)  \nonumber \\
& - & 1.306 - n_f ( -0.199 + 0.516 m_q - 0.421 m_q^2))
\alpha_s( \overline{m_b})^2
\label{eq:BaliBoyle}
\end{eqnarray}
%%%
%%%
where we have specialised to $c_{SW} = 1$.
%%%
Equation~\ref{eq:deltaMLOw} is a combination of the 
two loop static self energy and a conversion from
the bare coupling to the massless $\overline{MS}$
scheme~\cite{Martinelli:1998vt}. As stressed by 
Martinelli and Sachrajda~\cite{Martinelli:1998vt},
it is important to use a consistent coupling in
equation~\ref{eq:masterEq}, so that the poorly
behaved perturbative expansion of $Z_{pm}$ cancels
with that of $\delta m$. This makes it easier to use
the massless quark $\overline{MS}$ scheme.
The determination of $\Lambda_{QCD}$ 
includes the effects
of the masses of the 
light quarks~\cite{Booth:2001qp}. 
The use of~\ref{eq:BaliBoyle} changes the central value of 
$m_b(m_b)$ by 1 MeV for the data set DF5.

For our final result we use the central value from
data set $DF5$. The systematic uncertainties have been
discussed in this section. Hence our final result is
%%%
\begin{equation}
\overline{m_b} (\overline{m_b} ) = 
(4.25  \pm 0.02  \pm 0.03  \pm 0.03 \pm 0.08 \pm 0.06  ) {\mathrm GeV}
\end{equation}
%%%
%%%
%%%
where the errors are (from left to right):
statistical, perturbative, and neglect
of $1/m_b$ terms, ambiguities in the choice of
lattice spacing, and error in the choice of the mass
of the strange quark.

Gimenez et al.~\cite{Gimenez:2000cj} obtain
%%%
\begin{equation}
\overline{m_b} (\overline{m_b} ) = 
(4.26  \pm 0.03  \pm 0.05  \pm 0.07 ) {\mathrm GeV}
\end{equation}
%%%
from a simulation at $\beta = 5.6$ and volume = $24^3 \;40$
with two dynamical quark masses, from the $T\chi L $ collaboration.
The first error is due to statistics. The second error
includes the neglect of the $1/m_b$ terms and the ambiguity 
in the determination of the lattice spacing. The third error
is due to the neglect of higher order corrections
in the perturbative matching. Gimenez et al.~\cite{Gimenez:2000cj}
used a preliminary result for $\overline{X_2}$,
that was relatively imprecise~\cite{Burgio:1999ba} hence their
estimate of the higher order effects is looser than ours.
Also we used $\Lambda_{QCD}$ determined consistently
from this data set, while Gimenez et al.~\cite{Gimenez:2000cj} 
used continuum based estimates of the coupling.
This analysis has recently been updated by Di Renzo and
Scorzato~\cite{DiRenzo:2004xn} with the unquenched value of 
$X_2$.
Gimenez et al.~\cite{Gimenez:2000cj} only used 
one quantity to estimate the lattice spacing, so 
their error from the ambiguity in the choice of lattice spacing
is underestimated
in the final result. However, the two effects compensate and 
the final error is probably representative.
It is pleasing that our calculation with a different 
set of parameters is essentially consistent with 
that of Gimenez et al.~\cite{Gimenez:2000cj}.

\section{Conclusions} \label{eq:conc}

In table~\ref{tab:mbSummary} we collect some recent results for the
mass of the bottom quark from lattice QCD. Our result is consistent
with the previous unquenched calculations. Unfortunately, we have not
managed to reduce the size of the error bars.  The largest error in
the recent values for the mass of the bottom quark is due to the 
spread in different lattice spacings.  
Heitger and Sommer~\cite{Heitger:2003nj} noted that a change
in $r_0$ by 10\% changed the value of $m_b(m_b)$ by 150 MeV.

\begin{table}[tb]
  \centering
  \begin{tabular}{|c|c|c|} \hline
Group  &  comment & $m_b(m_b)$ GeV   \\ \hline
This work & unquenched & $4.25(2)(11)$  \\
Collins~\cite{Collins:2000sb} & unquenched & $4.34(7)_{-7}^{+0}$  \\
Gimenez et al.~\cite{Gimenez:2000cj} & unquenched & $4.26(9)$  \\
Di Renzo and Scorzato~\cite{Gimenez:2000cj} & unquenched & $4.21 \pm
0.03 \pm 0.05 \pm 0.04$  \\
Bali and Pineda~\cite{Bali:2003jq} & quenched & $4.19(6)(15)$  \\
Heitger and Sommer~\cite{Heitger:2003nj} & quenched & $4.12(8)$  \\
\hline
  \end{tabular}
  \caption{Lattice QCD results for $m_b(m_b)$ 
in the $\overline{MS}$ scheme. The last error on the 
Bali and Pineda result is an estimate of unquenching.}
\label{tab:mbSummary}
\end{table}
%%%

The prospects for an improved estimate of the mass of the bottom quark
from lattice QCD are quite good. The unquenched calculations with 
improved staggered quarks produce consistent lattice
spacings from many different quantities~\cite{Davies:2003ik}. 
The numerical calculation of the third order contribution
to $\delta m$ has been done for unquenched
Wilson fermions~\cite{DiRenzo:2004xn}.
Applying the technology of Heitger and Sommer~\cite{Heitger:2003nj}
to unquenched calculations would allow a non-perturbative 
estimate of the mass of the bottom quark that is free from 
problems with delicate cancellations of poorly converging
perturbative expressions. The use of automated perturbative 
calculations may allow the computation of the bottom
quark mass from NRQCD / FNAL type calculations with
two loop accuracy~\cite{Trottier:2003bw,Nobes:2003nc}.
We are investigating the use of the static formulation
introduced by the ALPHA collaboration~\cite{DellaMorte:2003mn},
but the required perturbative (or non-perturbative) factors
are not available yet.

Some combination of the techniques and projects mentioned in the last
paragraph should be able to produce a number of independent
calculations, with different systematic errors, of the mass of the
bottom quark from unquenched lattice QCD.

\section{Acknowledgements}

The lattice data was generated on the Cray T3E system at EPCC
supported by, EPSRC grant GR/K41663, PPARC grants GR/L22744 and
PPA/G/S/1998/00777. We are grateful to the ULgrid project of the
University of Liverpool for computer time.

%%\bibliographystyle{h-physrev2}
%%\bibliography{q_mass} 

\end{document}